\documentclass[final,3p,times,twocolumn]{elsarticle}

\usepackage{stmaryrd}
\usepackage{graphicx}
\usepackage{epstopdf}
\usepackage{verbatim}
\usepackage{amssymb}
\usepackage{amsmath}
\usepackage{comment}
\usepackage{color}
\usepackage{wasysym}

\journal{Computer Physics Communications}

\begin{document}

\def\nth{{$n^{\text{th}}~$}}

\begin{frontmatter}
% Title Page
\title{A Short Introduction to \\Numerical Linked-Cluster Expansions}
\author{Baoming Tang, Ehsan Khatami, and Marcos Rigol}
\address{Department of Physics, Georgetown University, Washington DC, 20057, USA}
\address{Physics Department, The Pennsylvania State University, 104 Davey Laboratory, 
University Park, Pennsylvania 16802, USA}

\begin{abstract}
We provide a pedagogical introduction to numerical linked-cluster expansions (NLCEs). 
We sketch the algorithm for generic Hamiltonians that only connect nearest-neighbor sites 
in a finite cluster with open boundary conditions. We then compare results for a specific model, 
the Heisenberg model, in each order of the NLCE with the ones for the finite cluster calculated 
directly by means of full exact diagonalization. We discuss how to reduce the computational
cost of the NLCE calculations by taking into account symmetries and topologies of the linked 
clusters. Finally, we generalize the algorithm to the thermodynamic limit, and discuss several 
numerical resummation techniques that can be used to accelerate the convergence of the series.
\end{abstract}

\begin{keyword}
Linked-cluster expansions, Exact diagonalization, Spin systems, Lattice models
\end{keyword}

\end{frontmatter}

\section{Introduction}

\subsection{High-temperature expansions}

Calculating exact finite-temperature properties of quantum lattice models can
be very challenging. One general approach to achieve that goal is to devise 
series expansions for the lattice model in question in the 
thermodynamic limit~\cite{HTE,linked}. A common example of these series expansions
are high-temperature expansions (HTEs), in which the partition function 
$\mathcal{Z}$ and other extensive properties of the system are expanded 
in powers of the inverse temperature $\beta=(k_BT)^{-1}$ (in what follows we set the 
Boltzman constant $k_B$ to unity). For example, consider the
Ising Hamiltonian with nearest-neighbor interactions:
\begin{equation}
\label{ising}
\hat{H}=-J\sum_{\left<i,j\right>} \sigma_i \sigma_j,
\end{equation}
where $J$ is the strength of the interaction, $\left<..\right>$ denotes 
nearest neighbors, and $\sigma_i~(=\pm1)$ is the Ising
spin on site $i$. The partition function can be written as
\begin{equation}
 \mathcal{Z}=\sum_{\{\sigma\}} e^{-\beta 
\hat{H}}=\sum_{\{\sigma\}} e^{\beta J \sum_{\left<i,j\right>}
\sigma_i \sigma_j},
\label{eq:z}
\end{equation}
where the sum runs over all possible configurations of the spins.
We define $K=\beta J$, which serves as a small parameter in the expansion:
\begin{equation}
\label{z2}
 \mathcal{Z}=\sum_{\{\sigma\}} \prod_{\left<i,j\right>} e^{K
\sigma_i \sigma_j}=\sum_{\{\sigma\}} \prod_{\left<i,j\right>} \sum_{l=0}^{\infty}
\frac{K^l}{l !} (\sigma_i \sigma_j)^l
\end{equation}
If we associate $(\sigma_i \sigma_j)^l$ to an $l$-fold bond between sites
$i$ and $j$, a typical term in the above expansion can be represented 
graphically as the lattice with various number of lines (including no 
line) connecting every two nearest-neighbor sites. Therefore, 
one can write the partition function in terms of contributions from all
possible graphs, up to a certain size, that can be embedded on the 
lattice. (In the Ising model, the fact that $\sigma=\pm1$ makes 
the calculations much easier than in quantum models such as the Heisenberg 
model.) The order in $K$ that each graph contributes to is equal to 
the number of bonds it has (see Ref.~\cite{linked}
for more details on this derivation). Equation~\eqref{z2} implies that the 
series converges only at high temperatures, above or of the order of $J$.
In what follows, we will see how this type of expansion is
related to linked-cluster expansions.

\subsection{Low-temperature expansions}

Similar to HTEs, low-temperature expansions (LTEs) can be devised to describe 
properties of a system with discrete excited states close to its ground
state, i.e., for $\beta\rightarrow\infty$. In this approach, the partition function is written as 
\begin{equation}
\mathcal{Z}=e^{-\beta E_{0}} \left[ 1+\sum_{n \neq 0} 
e^{-\beta(E_{n}-E_{0})}\right],
\end{equation}
where $E_0$ ($E_n$) is the ground state ($n^{\text{th}}$ excited state).
If there is an energy increment, $\epsilon$, satisfying $E_{n}-E_{0}=m_n\epsilon$
with $m_n$ being integer for all $n$, any thermodynamic property 
can be expressed as an expansion in powers of $e^{-\beta \epsilon}$. Then,  
cluster expansions similar to the ones discussed above for HTEs follow~\cite{linked}.

\subsection{Linked-cluster expansions}
\label{sec:lce}

The idea behind linked-cluster expansions (LCEs)~\cite{linked,sykes} is that for any extensive property $P$ 
of a lattice model (such as the logarithm of the partition function or the internal energy) 
one can compute its value per lattice site $P(\mathcal{L})/N$ in the thermodynamic limit
in terms of a sum over contributions from all clusters $c$ that can be embedded 
on the lattice:
\begin{equation}
\label{NLCE1}
P(\mathcal{L})/N=\sum_{c}L(c)\times W_{P}(c),
\end{equation}
where $L(c)$ is the multiplicity of $c$, namely, the number of ways per site in which
cluster $c$ can be embedded on the lattice, and $W_{P}(c)$ is the 
weight of that cluster for the property $P$. $W_{P}(c)$ is defined according to the 
inclusion-exclusion principle:
\begin{equation}
\label{NLCE2}
 W_{P}(c)=P(c)-\sum_{s \subset c} W_{P}(s),
\end{equation}
where 
\begin{equation}
\label{eq:p}
P(c)=\frac{\textrm{Tr} \left[\hat{P}(c)e^{-\beta \hat{H_c}}\right]} {\textrm{Tr} \left[e^{-\beta \hat{H_c}}\right]}
\end{equation}
is the property calculated for the finite cluster $c$ and 
the sum on $s$ runs over all sub-clusters of $c$. In Eq.~\eqref{eq:p}, $\hat{H_c}$ 
is the Hamiltonian of cluster $c$. One can check that the 
weight of a disconnected cluster vanishes because $P(c)$ can be written as
the sum of its parts (see Sec.~\ref{subsec:connect}), hence, 
the name linked-cluster expansions.

The convergence of the series in Eq.~\eqref{NLCE1}, when all the terms are 
considered, is guaranteed by the definition of weights in Eq.~\eqref{NLCE2}.
In fact, by swapping the place of $P(c)$ and $W_{P}(c)$ in Eq.~\eqref{NLCE2}, 
one can write the property of a {\em finite} cluster $c$ as the sum of 
its weight and the weights of its sub-clusters. Taking the thermodynamic
limit $c \to \mathcal{L}$ brings one back to Eq.~\eqref{NLCE1}. However, in that
limit, only a finite number of terms can be accounted for, and the series 
has to be truncated. 

Because of the inclusion-exclusion principle [Eq.~\eqref{NLCE2}], 
the weight of every cluster contains only the contribution to the property 
that results from correlations that involve all the sites in the cluster, 
and in a unique fashion in accord with its specific geometry. At low 
temperature, when correlations grow beyond the size of the largest 
clusters considered in the series, the results show a divergent behavior 
(due to the missing contributions of clusters in higher orders of the expansion). 
In most of the two-dimensional quantum models of interest, this occurs 
near or at zero temperature, e.g., for the nearest-neighbor 
antiferromagnetic (AF) Heisenberg model on a bipartite lattice.

The clusters in the sum \eqref{NLCE1} are usually grouped together based 
on common characteristics to represent different 
orders~\cite{M_rigol_06,m_rigol_07a}. For instance, in the {\em site expansion}, 
where sites are used as building blocks to generate the clusters, 
the order of the expansion is determined by the number of sites of the 
largest clusters. All the clusters with $n$ sites are said to
belong to the $n^{\text{th}}$ order. In LCEs, one has the freedom to 
devise an expansion (with a certain building block for generating the 
clusters in different orders) that best suites the particular model of 
interest. Some of these include the site, bond, or square expansions on 
the square lattice, and site, bond, or triangle expansions on the triangular and Kagome 
lattices, and so on~\cite{m_rigol_07a}.

Despite the simple form of the LCE equations above, its computational 
implementation can be challenging, as one has to (i) generate all the 
linked clusters that can be embedded on the lattice, (ii) identify their 
symmetries and topologies to compute multiplicities [this step dramatically 
reduces the computational effort as, for any given model, many 
different clusters have identical values of $P(c)$], (iii) identify the 
sub-clusters of each cluster to calculate weights, and (iv)
calculate the property of each cluster and perform the sums. LCEs 
are computationally very demanding as the number of embedded clusters, 
and their sub-clusters, grow exponentially with increasing the order 
of the expansion. Below, we explain all those steps in the context 
of an example (the site expansion on a finite square lattice).
We should stress that the HTE explained above for the Ising model can be 
seen as a LCE in which the property for each cluster is calculated using a 
perturbative expansion of Eq.~\eqref{eq:p} in terms of $K$.

\subsection{Numerical linked-cluster expansions}
\label{sec:nlce}

In this work, we present a pedagogical overview of the {\em numerical}
linked-cluster expansions (NLCEs) introduced in Ref.~\cite{M_rigol_06}. NLCEs 
use the basis of LCEs, but employ exact diagonalization (ED), instead of 
perturbation theory as done in HTEs, to calculate the properties of finite 
clusters in the series. This means that the properties of each cluster are calculated 
to all orders in $\beta$. The main advantage of NLCEs over HTEs is that, for models 
with short-range correlations, it is possible to access arbitrarily low temperatures 
because of the lack of a small parameter in the series. Furthermore,
for models in which correlations grow slowly as the temperature is lowered,
NLCEs can converge well below the temperature at which HTEs diverge.
In NLCEs, as opposed to HTEs, the convergence temperature is controlled by the 
correlations in the model, and by the highest order 
in the series that can be calculated.

The basics of NLCEs, and results for various spin and itinerant models 
in the square, triangular and kagome lattices, were presented in 
Refs.~\cite{M_rigol_06,m_rigol_07a,m_rigol_07b}. Recent applications
of this method exploring properties of frustrated magnetic systems
can be found in Refs.~\cite{singh_oitmaa_1,singh_oitmaa_2,applegate}. NLCE studies of the
Hubbard model in the square and honeycomb lattices were reported in 
Refs.~\cite{E_khatami_11b,khatami_rigol_12,tang_paiva_12}. How to deal with 
Hamiltonians and observables that break some of the symmetries
of the underlying lattice was discussed in 
Refs.~\cite{E_khatami_11,khatami_helton_12_69}. Finally, how to generalize
NLCEs to calculate ground-state as well as low-temperature properties
of lattice Hamiltonians with dimerized ground states was the subject
of Ref.~\cite{khatami_singh_11_66}, while direct ground state calculations
in the thermodynamic limit were done in Ref.~\cite{irving} and dynamics 
were explored in Ref.~\cite{dynamics}.

Here, we introduce NLCEs for finite clusters, in order to show how they converge 
to the exact result with increasing the order in the expansion, and, later,
discuss NLCEs in the thermodynamic limit. The exposition is organized as follows. 
In Sec.~\ref{sec:NLCE}, we present the algorithmic details and the 
numerical implementation of NLCE in two dimensions for a finite $4\times 4$ 
lattice. In Sec.~\ref{sec:res}, we report results 
of the expansion for the AF Heisenberg model on this cluster and compare 
them, in each order, to exact results that can be obtained by means of full 
exact diagonalization of the $4\times 4$ lattice. In Sec.~\ref{sec:thermo}, we 
discuss how to extend NLCEs to the thermodynamic limit, and compare NLCE 
results for the Heisenberg model to those from large-scale quantum Monte 
Carlo (QMC) simulations. In Sec.~\ref{sec:resum}, we describe two resummation
techniques that have been found to be widely applicable to accelerate 
the convergence of the NLCEs for thermodynamic properties 
of lattice models of interest.

\section{Implementation of NLCEs for Finite Systems}
\label{sec:NLCE}

\begin{figure}[t]
\centerline {\includegraphics*[width=1.2in]{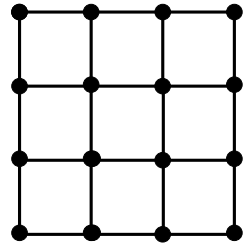}} 
\caption{The $4\times 4$ finite lattice with open boundary conditions used as an 
example in the derivation of the NLCE.}
\label{fig:lattice16}
\end{figure}

In this section, we sketch the NLCE algorithm for an arbitrary Hamiltonian that 
only connects nearest-neighbor sites on a $N=16$ site square lattice 
with open boundary conditions, which is shown in Fig.~\ref{fig:lattice16}. We 
have chosen this small lattice because it allows us to carry out the NLCE 
to all orders in the site expansion. It also allows us to compare the properties 
in each order of the expansion to exact results calculated directly by ED of 
the entire 16-site system.

\subsection{Embedded clusters}
\label{subsec:gener}

\begin{table}[!b]
\caption{Total number of embedded clusters (second column), number of linked-clusters (third column), 
number of linked-clusters that are not related by point-group symmetries (fourth column), and 
number of linked-clusters that are topologically distinct (fifth column), in each order of the site 
expansion for a 16-site lattice with nearest-neighbor interactions (Fig.~\ref{fig:lattice16}).}

\vspace{0.05in}

\begin{tabular}{|c|c|c|c|c|}
\hline
\color{blue} $n$ & \color{blue} $\frac{N!}{n!(N-n)!}$ & \color{blue}Connected &
\color{blue}Sym. distinct & \color{blue} Topo. \\
\hline
\color{blue}1 & 16 & 16 & 1 & 1 \\
\color{blue}2 & 120 & 24 & 1 & 1  \\
\color{blue}3 & 560 & 52 & 2 & 1 \\
\color{blue}4 & 1820 & 113 & 5 & 3 \\
\color{blue}5 & 4368 & 244 & 14 & 4 \\
\color{blue}6 & 8008 & 496 & 43 & 10 \\
\color{blue}7 & 11440 & 912 & 94 & 19 \\
\color{blue}8 & 12870 & 1474 & 197 & 49 \\
\color{blue}9 & 11440 & 2032 & 296 & 92  \\
\color{blue}10 & 8008 & 2286 & 327 & 167 \\
\color{blue}11 & 4368 & 2000 & 265 & 190 \\
\color{blue}12 & 1820 & 1236 & 169 & 152 \\
\color{blue}13 & 560 & 488 & 66 & 65 \\
\color{blue}14 & 120 & 116 & 20 & 20  \\
\color{blue}15 & 16  & 16 & 3 & 3 \\
\color{blue}16 & 1   & 1 & 1 & 1 \\
\hline
\end{tabular}
\label{tab:generation}
\end{table}

There exists $\binom{N}{n}=\frac{N!}{n!(N-n)!}$ clusters in the \nth order
on a lattice with $N$ sites. To identify them in a computer, one can 
number the lattice sites in an arbitrary fashion. Then, construct an array 
of size $N$ whose $i^{\textrm{th}}$ element is a binary number representing 
site number $i$ on the lattice. A 0 as an element of this array indicates that 
the corresponding site is not part of the generated cluster whereas a 1
indicates that it is part of the cluster. Therefore, all clusters in 
the \nth order can be generated by exploring all possible configurations of 
1 and 0 as elements of the above array, while keeping the total number of 
nonzero elements fixed at $n$. 

In the second column of Table~\ref{tab:generation}, 
we list $\binom{N}{n}$ for our example of the 16-site lattice in 
Fig.~\ref{fig:lattice16}. Once we have all the site clusters, we need to 
connect the sites present on them with bonds. This is done following the 
Hamiltonian of interest. For models with nearest-neighbor interactions, the 
case of interest here, bonds are added to every pair of nearest-neighbor sites 
\cite{M_rigol_06}. Hence, each site in our clusters is connected 
by bonds with up to 4 other sites. Examples of clusters within the site expansion 
are given in Fig.~\ref{fig:generation}.

\subsection{Connected clusters}
\label{subsec:connect}

Many of the site clusters generated in the step above contain disconnected parts.
Those clusters should be discarded because their weight $W_{P}(c)$, as given by 
Eq.~\eqref{NLCE2}, is zero. This can be easily seen if one assumes that cluster 
$c$ consists of two disconnected sub-clusters $c_{1}$ and $c_{2}$. Then, since
$P(c)$ is extensive, it can be written as the sum of the properties of the two 
sub-clusters:
\begin{equation}
\label{start1}
 P(c)=P(c_{1})+P(c_{2}).
\end{equation}
But, in addition to all of their subclusters, $c_1$ and $c_2$ themselves
are among the subclusters of $c$. Therefore,
\begin{eqnarray}
W_{P}(c)&=& P(c)-\sum_{s \subset c} W_{P}(s)\nonumber \\
&=&P(c)- \left[W_{P}(c_{1})+\sum_{s \subset c_{1}} 
W_{P}(s)\right]\nonumber \\
&\ &\ \ \ \ \ \ \ \ -\left[W_{P}(c_{2})+\sum_{s \subset c_{2}} W_{P}(s)
\right]\nonumber\\ &=&P(c)-P(c_{1})-P(c_{2})=0.
\end{eqnarray}

\begin{figure}[!b]
\centerline {\includegraphics*[width=3.in]{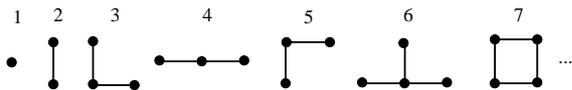}} 
\caption{A sample of linked clusters that can be embedded on our finite 16-site 
lattice. Clusters 3 and 5 are related by point-group symmetry and 
are topologically the same as cluster 4.}
\label{fig:generation}
\end{figure}

One then needs to find a way to distinguish between connected and disconnected 
clusters. A cluster is connected if starting from any site
one can reach any other site moving along the bonds (only nearest-neighbor bonds 
in our case). We have implemented this idea by reconstructing the 
cluster from scratch. In our codes, starting from one of the sites in
the cluster, we add bonds between that site and all of its neighboring 
sites, provided that they are part of the cluster. 
The same process is repeated for each of the new sites and this continues 
until there are no more options for adding bonds. The resulting cluster 
is then compared to the original one. If they are not identical, 
the cluster must have disconnected parts and is discarded.

In the third column in Table~\ref{tab:generation}, we show the number of 
clusters in each order of the site expansion after the disconnected ones 
have been removed. Figure~\ref{fig:generation}, depicts some of the linked 
clusters generated for our 16-site lattice.

\subsection{Point-group symmetries}
\label{subsec:symm}

If the Hamiltonian preserves the symmetries of the underlying lattice,
clusters related by point-group symmetries have the same weight 
for all observables. Therefore, they can be grouped together in order 
to avoid repeating the calculation of the weights for all of them. 
Depending on the lattice, one may have different number 
of point-group symmetries. For the square lattice, which is the case in 
our example, one has the following eight point group symmetries:
\\

\begin{tabular}{lll}
$\bullet$ Identity & \ \ & $\bullet$ Reflection about $x=0$\\
$\bullet$ Rotation by 90$^\circ$ & \ & $\bullet$ Reflection about $y=0$\\
$\bullet$ Rotation by 180$^\circ$ &\  & $\bullet$ Reflection about $x=y$ \\
$\bullet$ Rotation by 270$^\circ$ & \ & $\bullet$ Reflection about $x=-y$ 
\end{tabular}
\vspace{0.2in}

Hence, at this point in the implementation of the NLCEs, the goal is to 
identify and keep only those clusters that are not related by point-group 
symmetries. We achieve this through the following steps, which need to be 
followed independently for each order of the expansion:
\begin{itemize}

\item[(i)] Create a list of symmetrically distinct clusters and their 
multiplicity. This list begins with the first linked cluster in the 
order under consideration with multiplicity one.

\item[(ii)] Take the next connected cluster and generate all clusters that
are symmetrically related to it by applying the symmetry operations 
mentioned above.

\item[(iii)] For each cluster generated in (ii), check whether 
it is present in the list created in (i). This is achieved by finding out 
whether there is a translation vector that converts one cluster into the other.
If yes, increase the multiplicity of the stored cluster that is a match 
by one, and go to (ii).

\item[(iv)] If none of the clusters generated in (ii) is in the list created 
in (i), store the connected cluster in the list, set its multiplicity to 
one, and go to (ii).

\end{itemize}

This step significantly reduces the number of clusters that one has to consider. 
Among the clusters shown in Fig.~\ref{fig:generation}, clusters 3 and 5 are 
related by a 90$^\circ$ rotation and only one of them needs to be stored. 
The fourth column in Table \ref{tab:generation} shows the number of clusters that 
are not related by point-group symmetries in each order of the site expansion.

\subsection{Topological clusters}
\label{subsec:topol}

Furthermore, even if some clusters are not related by point-group symmetries, 
their Hamiltonian may be identical. For example, in models with only 
nearest-neighbor terms, clusters 3 and 4 in Fig.~\ref{fig:generation} 
have the same Hamiltonian. We then say that these two clusters are topologically 
equivalent, and only one of them needs to be diagonalized. It is important to note 
that topologically equivalent clusters share the same thermodynamic properties, 
but they may lead to different results for correlation functions. For example,
the distance between the two extreme sites in clusters 3 and 4 in 
Fig.~\ref{fig:generation} is $2a$ and $\sqrt{2}a$ ($a$ is the lattice spacing), 
respectively, and that difference needs to be taken into account when calculating 
correlation functions. Still, since the full exact diagonalization of each cluster 
is the most time consuming part of the NLCE calculations, the fact that one only 
needs to diagonalize topologically different clusters (or simply, 
{\em topological clusters}) reduces the computation time dramatically.

At this point, we need to generate a list of all 
topological clusters. It can be easily verified that they share the same 
topologically distinct sub-clusters, i.e., only diagonalizing topological 
clusters allows one to carry out the entire NLCE calculations.

The identification of the cluster topologies can be done through adjacency 
matrices ($M$). An adjacency matrix contains all the information about
the spatial relations between sites in the cluster. In the simplest 
form, an adjacency matrix of an $n$-site cluster for a model with only 
nearest-neighbors interaction can be a $n\times n$ matrix whose 
rows/columns represent different sites and the matrix elements, 
$M_{ij}$, are either 1, if sites $i$ and $j$ are nearest neighbors 
(are connected by a bond), or 0 otherwise. Such matrix will be symmetric 
with zeros as diagonal elements. A generalized version of such matrices 
can be used for models with bond anisotropy or correlations beyond 
nearest neighbors. 

If two clusters of the same size are topologically equivalent, then there 
exists a permutation of sites that transforms the adjacency matrix of one 
onto the other. Therefore, similar to the approach taken in the previous step, 
we make a list of topologically distinct clusters for each order as follows:

\begin{itemize}

\item[(i)] Create a list of topological clusters and their 
multiplicity. This list begins with the first symmetrically 
distinct cluster in the order under consideration with its 
multiplicity.

\item[(ii)] Take the next symmetrically distinct cluster and 
construct its $n!$ possible adjacency matrices (given the 
$n!$ site permutations).

\item[(iii)] For each adjacency matrix generated in (ii), check whether 
it coincides with the adjacency matrix of any of the clusters in the 
list created in (i). If yes, increase the multiplicity of the stored 
cluster that is a match by the multiplicity of the new cluster, 
and go to (ii).

\item[(iv)] If none of the adjacency matrices generated in (ii) 
matches the one of a cluster in the list created in (i), store 
the newly found topological cluster in the list, set its multiplicity 
to the one it had as a symmetrically distinct cluster, and go to (ii).

\end{itemize}

\begin{figure*}[ht]
\parbox[l]{2in}{
\centerline {\includegraphics*[width=0.6in]{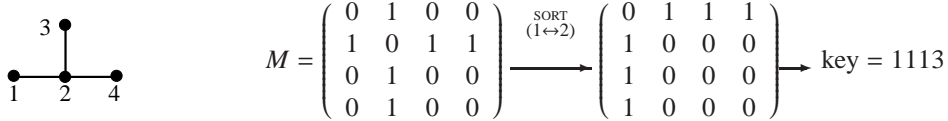}}}
\parbox[r]{3.5in}{
\begin{equation}
M=\left( \begin{array}{cccc}
       0 & 1 & 0 & 0  \\
       1 & 0 & 1 & 1  \\   
       0 & 1 & 0 & 0  \\
       0 & 1 & 0 & 0  \\
       \end{array}
\right) \stackrel{\stackrel{\rm SORT}{(1\leftrightarrow 2)}}{\vector(1,0){30}}
\left( \begin{array}{cccc}
       0 & 1 & 1 & 1  \\
       1 & 0 & 0 & 0  \\   
       1 & 0 & 0 & 0  \\
       1 & 0 & 0 & 0  \\
       \end{array}
\right) \vector(1,0){10} \ \ {\rm key}= 1113 \nonumber
\end{equation}}
\caption{An example of the adjacency matrix that represents a 4-site 
cluster (left). Rows and columns are representatives of the sites 
on the cluster. A 1 (0) as the $(i,j)$ element of the matrix indicates 
that sites $i$ and $j$ are connected (not connected) by a nearest-neighbor
bond. The matrix on the right is sorted by exchanging sites 1 and 2.
A key is associated to the sorted matrix according to an arbitrary 
formula [here, we use Eq.~\eqref{eq:key}].}
\label{fig:key}
\end{figure*}

In practice, this part of the algorithm can be very time consuming not 
only because the number of permutations can be very large, but also 
because of the matrix comparisons. So, alternatively, one can generate a 
{\em key} (an individualized number) for each adjacency matrix and use 
that key for comparison purposes. To do that, one has to pick only one, 
out of the $n!$ adjacency matrices to represent a cluster. This can be 
accomplished by labeling sites according to a unique 
criterion for the adjacency matrix. For example, if we think of each 
adjacency matrix as a large binary number by attaching its columns/rows 
one after another, one can pick the permutation of site numbers 
that yields the largest (or smallest) binary number. In Fig.~\ref{fig:key}, 
we show one of the adjacency matrices of a four-site cluster and the 
same matrix after exchanging site labels 1 and 2. The latter, which 
yields the smallest binary number that consists of columns of the matrix, 
is the one that will represent this cluster and that will be stored.

The recipe for generating the keys is arbitrary and, for large cluster
sizes, one may not be able to find a recipe that guarantees a unique key
for each cluster, i.e., multiple adjacency matrices may end up sharing 
the same key. The following is an example of one such recipe for creating 
the keys:
\begin{equation}
\label{eq:key}
\textrm{key}=\sum_{i,j} M_{ij} (i\times j)^{i+j}.
\end{equation}
If the key is not unique, then when the key of a new cluster matches the 
one of a stored topological cluster, one has to directly compare 
the adjacency matrices in order to determine whether the two clusters are 
topologically the same. We emphasize that the use of keys accelerates 
the process of comparing adjacency matrices tremendously and has been 
implemented in all our NLCEs.

Up to this point, we have identified and stored the topological clusters
of each size, their multiplicities, and their adjacency matrices with their
respective keys. The number of topological clusters in each order of our 
example is shown in the last column in Table.~\ref{tab:generation}. This table 
now makes apparent the significant reduction of the number of clusters that
need to be fully diagonalized from the initial $\binom{N}{n}$. In the following 
step, we explain how to enumerate the sub-clusters of each topological cluster 
to be able to ultimately calculate its weight.

We should stress that the approach described here to identify topologically 
distinct clusters works so long as the cluster sizes are small enough that 
all permutations of the introduced adjacency matrices can be explored. For 
larger cluster sizes, one needs to generate more sophisticated adjacency 
matrices whose description is beyond the scope of this  
introduction to NLCEs.
 
\subsection{Sub-clusters}
\label{subsec:subcluster}

Given the description so far for finding topological clusters, 
identifying all subclusters of a topological cluster is a relatively 
straightforward task. Note that finding the topological clusters of our 
finite 16-site lattice can, in itself, be interpreted as finding all 
subclusters of a $4\times 4$ cluster in the list of topological 
clusters of a larger lattice. Therefore, following the same procedure 
as in Sec.~\ref{subsec:gener} and Sec.~\ref{subsec:connect}, and replacing 
the $4\times 4$ system with any of the topological clusters, we first 
generate all possible connected subclusters. This time, there is no need 
for checking the point-group symmetries of the subclusters 
(Sec.~\ref{subsec:symm}). This is because all clusters that are related 
by symmetry share the same topology and have the same adjacency matrix. So, 
after generating a new connected subcluster, it is sufficient to construct 
its adjacency matrix and compare it to those of clusters with the same size 
that are stored in the list of topological clusters. Since 
there will always be a match, one only needs to keep track of how many 
subclusters of a certain topology a topological cluster has. The latter can be achieved 
by considering a multiplicity matrix, $Y$, whose $(i,j)$ element gives
the number of times the $j$th topological cluster can be embedded in the $i$th 
topological cluster.

The steps described from Sec.~\ref{subsec:gener} to Sec.~\ref{subsec:subcluster}
need to be carried out just once for all Hamiltonians involving only 
nearest-neighbor terms in the square lattice. The table with all topological 
clusters and subclusters can be stored and used for different nearest 
neighbor Hamiltonians and, for any given Hamiltonian, for different 
microscopic parameters.

\subsection{Weights}
\label{subsec:weights}

Now, we have all the necessary tools to account for the subcluster
subtractions in order to calculate the weights in Eq.~\eqref{NLCE2}. 
The steps that follow need to be carried out every time that a new 
Hamiltonian or Hamiltonian parameter is explored. One starts with
the smallest cluster in the first order. That cluster has no subclusters 
and, therefore, the weight of the property $P$ in the cluster is simply 
equal to the property $P(1)$. Then, to compute the weight of 
the next topological cluster (a single bond in the second order of our 
site expansion, see Fig.~\ref{fig:generation}), we have to subtract 
the weight of the cluster in the first order from its property, 
according to the multiplicities given by the matrix $Y$:
\begin{eqnarray}
W_P(1) &=& P(1) \nonumber \\
W_P(2) &=& P(2) - Y_{21}W_P(1) = P(2) - Y_{21}P(1)\nonumber\\
\vdots
\end{eqnarray}
In the above equations, indices 1 and 2 refer to the cluster number, $c$, 
which is not necessarily the order number. As mentioned in Sec.~\ref{sec:lce}, 
in NLCEs, the property $P(c)$ is calculated using ED. 
For instance, the internal energy is:
\begin{equation}
\label{eq:E}
E(c)=\langle \hat{H_c} \rangle=\frac{\sum_{n=1}^{M_c}\varepsilon_{n}(c)
\exp[-\beta \varepsilon_{n}(c)]} {\sum_{n=1}^{M_c} \exp[-\beta 
\varepsilon_{n}(c)]},
\end{equation}
where $\hat{H_c}$ is the Hamiltonian for the finite cluster $c$ with 
eigenvalues $\varepsilon_n(c)$, and $M_c$ is the size of the Hilbert 
space on that cluster.

We define partial sums, $S$, to group together contributions from topological 
clusters with a common characteristic in one order of the NLCE. 
For example, in the site expansion, the most natural characteristic is the 
number of sites. Therefore, $S_n$, which is the sum of the contributions 
to property $P$ from all $n$-site topological clusters in the series 
($c_n$) reads
\begin{equation}
S_n = \sum_{c_n} L(c_n)W_P(c_n).
\end{equation}
The property in the $m$th order of NLCE is then
\begin{equation}
\label{eq:order}
P_m(\mathcal{L})/N = \sum_{n=1}^{m} S_n.
\end{equation}

\subsection{Results for the AF Heisenberg model}
\label{sec:res}

Now that we have sketched the site expansion for a 16-site lattice and for 
generic Hamiltonians that connect only nearest-neighbor sites, let us 
consider a specific example of one such Hamiltonians, namely, the
AF Heisenberg model
\begin{equation}
\label{eq:ham}
\hat{H}=J\sum_{\left<i,j\right>} \hat{\bf S}_{i} \cdot \hat{\bf S}_{j},
\end{equation}
where $\hat{\bf S}_i$ is the spin operator at site $i$. For simplicity, we set
$J=1$ so that, in what follows, all energies are given in units of $J$.

\begin{figure}[!t]
\centerline {\includegraphics[width=0.4\textwidth]{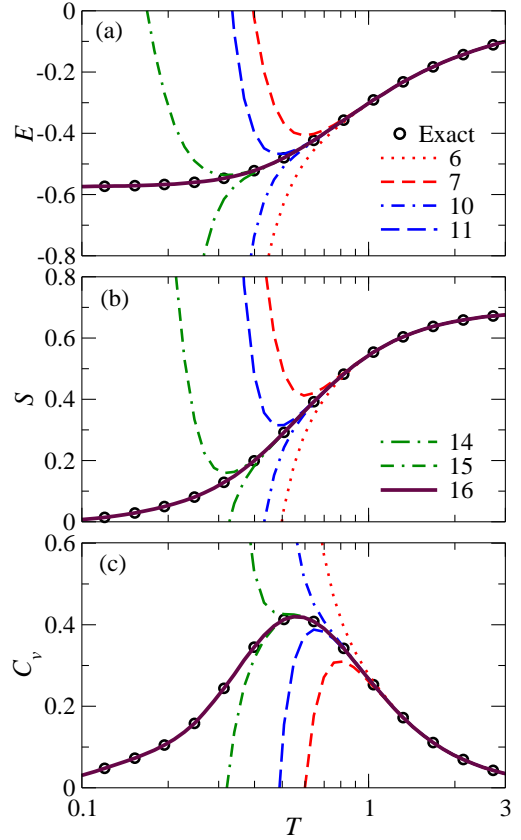}}
\caption{(a) Energy, (b) entropy, and (c) specific heat per site vs
temperature for the AF Heisenberg model on the 16-site cluster in
Fig.~\ref{fig:lattice16}. Exact results are obtained by full exact
diagonalization of the 16-site cluster with open boundary conditions and are
compared to those in different orders of the site expansion explained
throughout this work.}
\label{fig:finite}
\end{figure}

In Fig.~\ref{fig:finite}, we show results for the energy per site (a), the entropy per site (b),
\begin{equation}
S=\frac{1}{N}\left( \ln Z + 
\frac{ \left\langle{\hat{H}}\right\rangle}{T} 
\right),
\end{equation}
and the specific heat per site (c),
\begin{equation}
C_v= \frac{1}{N}\frac{\langle {\hat{H}}^2 \rangle-\langle {\hat{H}} \rangle^2}{T^2}, 
\end{equation}
on the 16-site lattice. The exact results (labeled ``Exact'' in Fig.~\ref{fig:finite}) were 
obtained by means of full diagonalization of the Hamiltonian, and are compared 
to those obtained in different orders of the site expansion [Eq.~\eqref{eq:order}]. 
Two things are apparent in those plots: (i) with increasing order, the NLCE results converge 
to the exact ones at lower temperatures. (ii) At any given order, quantities that can be 
represented by lower order derivatives of the partition function converge to lower temperature. 
Note the contrast 
between the results for the energy and $C_v$. Interestingly, even in the 15th order, 
when only the contribution from the 16-site cluster is missing, the energy has a 
visible discrepancy with the exact result at temperatures as high as $T=0.3$.

\section{Thermodynamic Limit}
\label{sec:thermo}

The generalization of the NLCE presented above to the 
thermodynamic limit is straightforward. Only certain steps in the algorithm 
change when dealing with the infinite-size system. In this section, we 
discuss those changes and show NLCE results for the Heisenberg model 
[Eq.~\eqref{eq:ham}] in the thermodynamic limit.

First, for the generation of the clusters, we start with the smallest one 
(a site in the site expansion) and systematically add more building blocks 
(sites in the site expansion). Hence, to generate clusters with $n+1$ sites, 
we consider all symmetrically distinct clusters that have $n$ sites and add 
a nearest-neighbor site to every site at the edge of the cluster. Note that 
this way of building site clusters works because of  translational 
invariance in the infinite lattice, and it is not appropriate for the finite 
system in Fig.~\ref{fig:lattice16} where not all sites are equivalent. 
This approach not only guarantees the generation of all possible clusters
that can be embedded on the infinite lattice, but also produces only connected 
clusters, i.e., the step presented in Sec.~\ref{subsec:connect} (examining 
the connectivity of the clusters) can be skipped. The second column in Table 
\ref{tab:generationTL} show the number of connected clusters per site, 
with up to 17 sites, 
in the site expansion of the square lattice in the thermodynamic limit.

For finite systems without translational symmetry, all the different
embeddings of a particular cluster are automatically
generated in the approach discussed in Sec.~\ref{subsec:gener}. 
Thus, the multiplicity of a symmetrically 
distinct cluster is calculated by counting the times it appears
in the process of generating all clusters. For the infinite system, 
because of translational symmetry, the multiplicity is simply equal 
to the number of point-group symmetries that transform a cluster to 
another cluster that is not related to the first one by a translation. 
Hence, a process similar to the one carried out for identifying 
the symmetries of the clusters on a finite lattice needs to be implemented 
for the clusters of the infinite system. However, in the latter, 
the multiplicity of symmetrically distinct clusters are determined immediately 
after building them by just examining their point-group symmetries. 
Moreover, if by adding sites to a cluster in the previous order one finds 
a new cluster that is related by a symmetry operation to one of the  
clusters stored in the new list, we simply dismiss it. The third column in 
Table \ref{tab:generationTL} show the number of symmetrically distinct clusters per site, 
with up to 17 sites, in the site expansion of the square lattice in the 
thermodynamic limit.

\begin{figure}[!t]
\centerline {\includegraphics[width=0.4\textwidth]{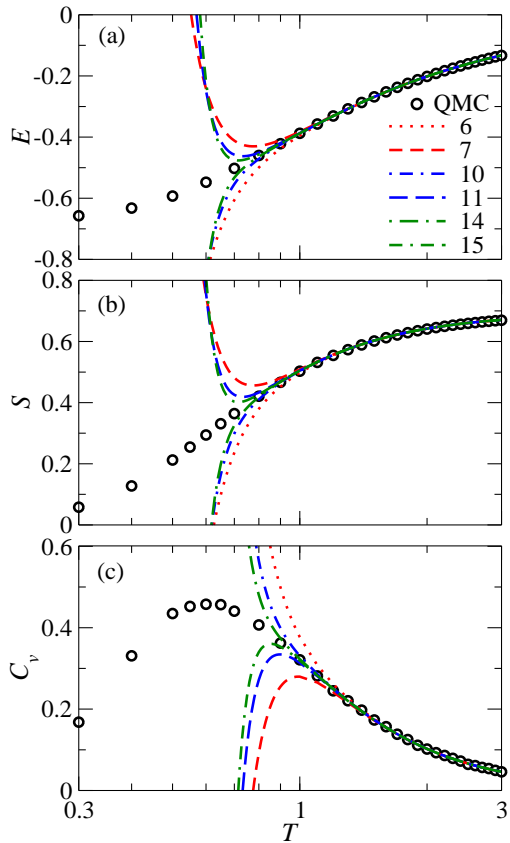}}
\caption{(a) Energy, (b) entropy, and (c) specific heat per site of the AF
Heisenberg model on the square lattice in the thermodynamic limit as a
function of temperature. NLCE bare sums up to 15th order are compared with QMC
results for a 256 $\times$ 256 lattice.}
\label{fig:E_Thermo_Bare}
\end{figure}

\begin{table}[!b]
\caption{Total number of linked-clusters (second column), 
number of linked-clusters that are not related by point-group symmetries (third column), and 
number of linked-clusters that are topologically distinct (fourth column) per site, up to the 17th order 
of the site expansion of the square lattice (in the thermodynamic limit) with nearest-neighbor 
interactions.}

\vspace{0.05in}

\begin{tabular}{|c|c|c|c|c|}
\hline
\color{blue} $n$ & \color{blue}Connected &
\color{blue}Sym. distinct & \color{blue} Topo. \\
\hline
\color{blue}1 & 1 & 1 & 1 \\
\color{blue}2 & 2 & 1 &  1  \\
\color{blue}3 & 6 & 2 &  1 \\
\color{blue}4 & 19 & 5 & 3 \\
\color{blue}5 & 63 & 12 & 4 \\
\color{blue}6 & 216 & 35 & 10 \\
\color{blue}7 & 760 & 108 & 19 \\
\color{blue}8 & 2725 & 369 & 51 \\
\color{blue}9 & 9910  & 1285 & 112 \\
\color{blue}10 & 36446  & 4655 & 300 \\
\color{blue}11 & 135268 & 17073 & 746 \\
\color{blue}12 & 505861 & 63600 & 2042 \\
\color{blue}13 & 1903890 & 238591 & 5450 \\
\color{blue}14 & 7204874 & 901971 & 15197  \\
\color{blue}15 & 27394666 & 3426576 & 42192 \\
\color{blue}16 & 104592937 & 13079255 & 119561 \\
\color{blue}17 & 400795844 & 50107909 & 339594 \\
\hline
\end{tabular}
\label{tab:generationTL}
\end{table}

After finding all symmetrically distinct clusters, we follow exactly the 
same procedure described in Secs.~\ref{subsec:topol} and \ref{subsec:subcluster} to determine all 
topological clusters and their subclusters. For the site expansion in the square lattice, the last 
column in Table \ref{tab:generationTL} show the number of topological clusters per site
in the thermodynamic limit. 
We should stress that all the steps described above need to be carried out 
just once for all lattice Hamiltonians that only connect nearest-neighbor 
sites in the square lattice.

Once the topological clusters and subclusters are determined, one can 
study any nearest-neighbor model of interest. In Fig.~\ref{fig:E_Thermo_Bare}, 
we show results for the energy (a), the entropy (b), and the specific heat (c)
per site for the AF Heisenberg model on the square lattice, and up to the 15th
order of the site expansion. NLCE results are compared to those obtained 
by means of QMC simulations, previously reported in Ref.~\cite{E_khatami_11}, 
for a very large periodic system with $256\times256$ sites. Those plots 
exhibit qualitatively similar features to the ones in Fig.~\ref{fig:finite}
for a finite cluster. Namely, as the order increases, NLCE results 
converge to lower temperature. In addition, the energy 
and the entropy can be seen to converge to lower temperature than
the specific heat. We note that, only in the 15th order, there are 42,192 
topological clusters that need to be diagonalized (see Table 
\ref{tab:generationTL}). Because of the very large number of clusters, 
one can use an embarrassingly parallel code that distributes groups of 
clusters to different processors so that one diagonalizes 
many of them at once in every order of the NLCEs.

\section{Resummation Algorithms} 
\label{sec:resum}

As seen from the results in Fig.~\ref{fig:E_Thermo_Bare}, the bare sums of the NLCE 
convergence down to a temperature that depends on the order up to which the expansion 
can be carried out and, on a more fundamental level, it depends on the build up of 
correlations in the system as the temperature is lowered. The longer the correlations 
the larger the cluster sizes that need to be included in the series to
achieve convergence. Fortunately, even if one cannot calculate more orders
of the NLCE, because of the exponential increase of the number of clusters and 
of the Hilbert space of each cluster, several numerical resummation algorithms 
have been developed that accelerate the convergence of NLCEs \cite{m_rigol_07a}.

The goal of those algorithms is to estimate 
$P(\mathcal{L})/N=\text{lim}_{m\rightarrow\infty}P_m$ in Eq.~\eqref{eq:order} from 
a finite set $\{P_m\}$. Resummation techniques then provide results for the observables of 
interest in regions where the bare sums [Eq.~\eqref{eq:order}] do not converge. 
Here we will focus our discussion on two such methods: 
Wynn's algorithm~\cite{Wynn} and Euler's transformation~\cite{Euler}.
They have been shown to be extremely useful in accelerating the convergence
of NLCEs for thermodynamic properties in several models of interest 
\cite{M_rigol_06,m_rigol_07a,m_rigol_07b,E_khatami_11b,khatami_rigol_12,tang_paiva_12,
E_khatami_11,khatami_helton_12_69,khatami_singh_11_66}. 

In Wynn's algorithm, one defines
\begin{eqnarray}
\label{eq:wynn}
 \varepsilon_{n}^{(-1)}=0,~~ \varepsilon_{n}^{(0)}=P_n, \nonumber \\ 
 \varepsilon_{n}^{(k)}=\varepsilon_{n+1}^{(k-2)}+\frac{1}{\Delta\varepsilon_{n}^{(k-1)}},  
\end{eqnarray}
where the differentiating operator $\Delta$ is applied to subscripts and 
is expressed as
\begin{equation}
  \Delta\varepsilon_{n}^{(k-1)}=\varepsilon_{n+1}^{(k-1)}-\varepsilon_{n}^{(k-1)}.
\end{equation}

\begin{figure}[!t]
\centerline {\includegraphics[width=0.4\textwidth]{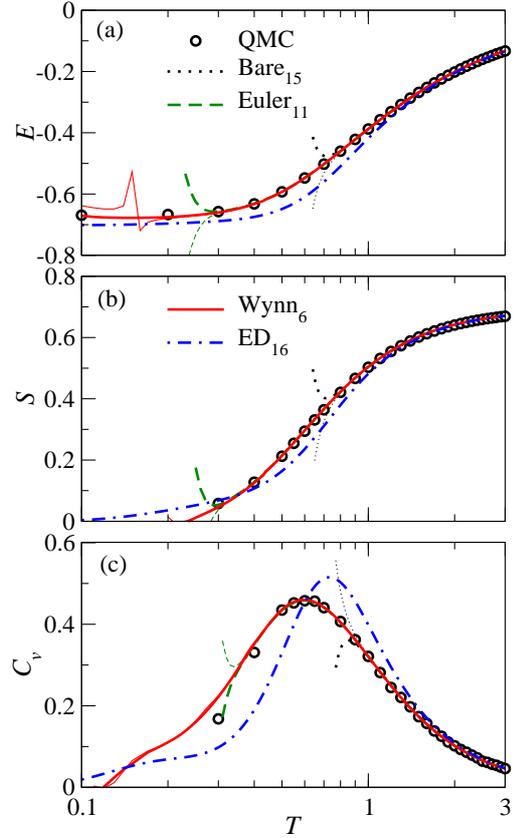}}
\caption{(a) Energy, (b) entropy, and (c) specific heat per site of the AF 
Heisenberg model on the square lattice as a function of temperature. Here we show results of 
last two orders (thick lines are used for the last order and thin lines for the one to last order) 
for the NLCE bare sums (up to the 15th order), Wynn's algorithm with 6 cycles of improvement, 
and Euler's transformation (for the last 11 orders). Results from a large-scale QMC and exact 
diagonalization of a 16-site cluster with periodic boundary conditions are also shown.}
\label{fig:Thermo_limit}
\end{figure}

It is expected that the even entries $\varepsilon_{n}^{(2l)}$ converge to 
$P(\mathcal{L})/N$, while the odd ones $\varepsilon_{n}^{(2l+1)}$ usually diverge,
where $l$ is defined as the number of cycles of improvement. As an example,
assume that the highest order in the NLCE that can be considered is $m$, i.e., the 
bare sum yields $P_m$ for the property in the last order. Now, assume that 
we use the Wynn algorithm with $l=1$ (one cycle of improvement) instead. 
In that case,
\begin{equation}
P_m\rightarrow \varepsilon_{m-2}^{(2)}=P_{m}-\frac{S_{m}}{1-\frac{S_{m-1}}{S_{m}}},
\end{equation}
where the second term in the right hand side is the first order correction to 
the result of the bare sum. Yet, it can often significantly improve the convergence.
Note that, for every cycle of improvement, there will be two terms less in 
the new series determined by means of Eq.~\eqref{eq:wynn} ($P_m\rightarrow \varepsilon_{m-2l}^{(2l)}$).

We have found that this algorithm is one of the best for accelerating the 
convergence of NLCEs. In Fig.~\ref{fig:Thermo_limit}(a), 
we show the same energy as in Fig.~\ref{fig:E_Thermo_Bare}(a) but include the results 
obtained after six cycles of improvement using Wynn's algorithm. One can see that the 
last two orders of the Wynn sum (red solid lines) agree with each other down
to much lower temperatures ($T\sim 0.2$) in comparison to the bare sums. In addition, 
the last order (thick solid line) is in perfect agreement with the QMC results in the entire 
temperature region shown. Very similar results can be seen for the entropy in 
Fig.~\ref{fig:Thermo_limit}(b). Furthermore, as depicted in Fig.~\ref{fig:Thermo_limit}(c), 
and unlike the bare sums, the Wynn algorithm can precisely capture the position 
and the height of the peak in the specific heat of this model, and even converges 
to temperatures lower than those accessible to the QMC calculations depicted 
in the figure. 

Note that the maximum cluster sizes considered in NLCE shown there 
(15 sites) are far smaller than the size of the lattice in the QMC simulations 
($256\times 265$). Hence, we see that a thorough exploration of all topologies in 
NLCEs, up to those cluster sizes, completely eliminates finite-size effects. In addition, 
in combination with resummation algorithms, one can compute observables up to quite 
low temperatures. NLCE results can also be contrasted against those obtained from ED of a 
16-site cluster with periodic boundary conditions. As seen in Fig.~\ref{fig:Thermo_limit},
the finite cluster results depart from QMC at temperatures higher than the 
convergence temperature of NLCE even with the bare sum. Using slightly larger 
clusters, which can still be solved by means of full exact diagonalization, 
does not improve this trend either~\cite{E_khatami_11}. 

We should point out that the AF Heisenberg model on the square lattice is one of the 
most challenging models that one can attempt to solve with NLCEs. This is because 
the antiferromagnetic correlations grow exponentially with decreasing temperature, 
and quickly exceed the linear size of the largest clusters that can be treated 
within ED. For this reason, frustrated magnetic systems, for which QMC techniques 
run into the infamous sign problem, and where correlation lengths are small 
and grow slowly with decreasing temperature, are ideal to be explored by means of
NLCEs.

Another very useful resummation technique for alternating series, i.e., 
series in which $S_n$ alternates in sign, is the Euler transformation. Within
this approach, $S_n$ is replaced by $u_n=(-1)^n S_n$ and $P(\mathcal{L})/N$ is 
approximated by the following sum:
\begin{equation} \label{Euler}
u_0-u_1+u_2+\dots-u_{n-1}+\sum_{l=0}^{m-n}\frac{(-1)^l}{2^{l+1}}\Delta^{l}u_{n},
\end{equation}
where $\Delta$ is defined as forward differencing operator
\begin{eqnarray}
\Delta^0 u_{n} &=& u_{n}, \nonumber \\ 
\Delta^1 u_{n} &=& u_{n+1} - u_{n}, \nonumber \\ 
\Delta^2 u_{n} &=& u_{n+2} - 2u_{n+1} + u_{n}, \nonumber \\
\Delta^3 u_{n} &=& u_{n+3} - 3u_{n+2} + 3u_{n+1} - u_{n}, \\
\vdots \nonumber
\end{eqnarray}
and $n-1$ is the number of terms for which a bare sum is performed
before the Euler transformation sets in for higher order terms. The first
few terms of the approximation can be written as
\begin{equation}
P_m\rightarrow P_{n-1}+(-1)^n\left[\frac{1}{2}S_{n}+\frac{1}{4}(S_{n}+S_{n+1})+\cdots\right].
\end{equation}

It is evident from the results in Fig.~\ref{fig:Thermo_limit} after implementing 
the Euler transformation for the last 11 terms ($n=5$), that the latter 
algorithm also dramatically improves the convergence from that of the bare sum.

\section*{Acknowledgments}

This research was supported by the National Science
Foundation (NSF) under Grant No. OCI-0904597 and enabled
by an allocation of advanced computing resources, supported
by the NSF. The computations in Sec.~\ref{sec:thermo}
were performed on Kraken at the National Institute for
Computational Science under Account No. TG-DMR100026.

\end{document}